\documentclass[twocolumn]{aastex63}

\def\a{\alpha}
\def\b{\beta}
\def\g{\gamma}
\def\y{\gamma}
\def\d{\delta}
\def\D{\Delta}
\def\e{\epsilon}
\def\i{\iota}
\def\k{\kappa}
\def\l{\lambda}
\def\u{\mu}
\def\v{\nu}
\def\s{\sigma}
\def\t{\tau}

\def\x{\xi}

\def\n{\nabla}
\def\w{\omega}
\def\z{\zeta}

\def\p{\phi}

\def\L{\Lambda}

\def\O{\Omega}
\usepackage{lipsum}

\shorttitle{\textsc{Follow-up of \textit{NuSTAR} J053449+2126.0}}
\shortauthors{Rodriguez et al.}
\graphicspath{{./}{figures/}}

\usepackage{float}
\usepackage{appendix}
\usepackage{xcolor}
\usepackage{amsmath}

\begin{document}

\title{The Search for a Counterpart to \textit{NuSTAR} J053449+2126.0}

\correspondingauthor{Antonio C. Rodriguez}
\email{acrodrig@caltech.edu}

\author{Antonio C. Rodriguez}
\affiliation{California Institute of Technology, Department of Astronomy \\
1200 East California Blvd\\
Pasadena, CA, 91125, USA}
\author{Yuhan Yao}
\affiliation{California Institute of Technology, Department of Astronomy \\
1200 East California Blvd\\
Pasadena, CA, 91125, USA}
\author{Kishalay De}
\affiliation{MIT-Kavli Institute for Astrophysics and Space Research\\ 77 Massachusetts Ave.\\ Cambridge, MA 02139, USA}

\author{S.R. Kulkarni}

\affiliation{California Institute of Technology, Department of Astronomy \\
1200 East California Blvd\\
Pasadena, CA, 91125, USA}

\begin{abstract}
\cite[][ATel \#15171]{2022ATel15171....1T} have recently reported the discovery of an X-ray source, \textit{NuSTAR} J053449+2126.0, during a calibration observation which took place on 25 April 2020. We scan the Zwicky Transient Facility (ZTF) alerts and archival photometry to determine the nature of the source. Palomar Gattini-IR is searched as well. We identify no obvious counterpart candidate. Follow-up X-ray and optical studies are needed to determine the true counterpart.
\end{abstract}

\section{Introduction}
The confirmation of an optical or infrared counterpart to an X-ray source is often needed to confirm the nature of the source. Accreting systems are among the most common X-ray sources: galactic sources include compact object binaries such as cataclysmic variables (CVs) and neutron star (NS)/black hole (BH) binaries \citep[e.g.][]{1983ARA&A..21...13B} while extragalactic sources are often active galactic nuclei (AGN) \citep[e.g.][]{2007ihea.book.....R}.

ATel \#15171 \citep{2022ATel15171....1T} has recently reported the discovery of an X-ray source, \textit{NuSTAR} J053449+2126.0, but it is unclear whether the source is an X-ray transient or persistent X-ray source from the ATel alone. The authors report the extraction of a 1 arcmin spectrum centered on the object positioned at (05:34:49.20, +21:26:02.9). The spectrum is fitted with a power-law with photon index $\sim 2.46$ in the 3.0--20.0 keV band. With a flux of $\sim 4 \times 10^{-13} \textrm{erg/s/cm}^2$ in the 3.0--10.0 keV band, assuming the source at $z = 0.1$ would put its X-ray luminosity at $\sim 5 \times 10^{42} \textrm{erg/s}$. Assuming it to be a galactic X-ray source at 1 kpc would put its X-ray luminosity at $\sim 3 \times 10^{31} \textrm{erg/s}$. The Galactic coordinates ($\ell$:185.0864077, $b$:-6.0382378) of this source are a few degrees away from the Crab nebula, leading us to speculate it may be a Galactic source.

\section{ZTF Search}
The Zwicky Transient Facility (ZTF) is a photometric survey that uses a wide 47 $\textrm{deg}^2$ field-of-view camera on the Samuel Oschin 48-inch telescope at Palomar Observatory with $g$, $r$, and $i$ filters \citep{bellm2019, graham2019, dekanyztf, masci_ztf}.  

We use ZTF Data Release 5 (DR5) with forced photometry provided for the ZTF collaboration through the Infrared Processing and Analysis Center\footnote{\url{https://irsa.ipac.caltech.edu/Missions/ztf.html}} \citep[IPAC;][]{masci_ztf}. The forced photometry calculates photometry of the object on all available frames by forcing the location of the PSF to remain fixed according to the ZTF absolute astrometric reference. 


\subsection{ZTF and Palomar Gattini-IR Alerts}
We first searched the ZTF Alerts catalog and found two sources within a 1 arcmin radius. Two alerts were discovered, ZTF21aaxwqyj and ZTF21aaxwtnr, which we attribute to the same source: an asteroid or other Solar System object. Both alerts appear only once and are consistent with a moving object.

The Palomar Gattini-IR \citep{de_gattini} infrared telescope alerts were also searched within a 1 arcmin radius of \textit{NuSTAR} J053449+2126.0, and one source was found to have some infrared variability, 2MASS J05344864+2125080. The ZTF light curve does not show any periodicity or signifcant outbursts.


\subsection{ZTF light curves}
We then downloaded all light curves within 1 arcmin of \textit{NuSTAR} J053449+2126.0 within the ZTF Data Release 5 (DR5) catalog. We found 202 light curves, searching between both $r$ and $g$ filters. The two criteria we searched for were either: 1) an outburst in the light curve or 2) a strong periodic signal indicative of a binary system. No ($>$1 mag) outbursts were found within a 2 day window of 25 April 2020.
Strong outbursts \textit{anywhere} in the light curves were then searched. Two objects with significant ($>$2 mag) outbursts in thir light curves were found, ZTFJ053452+212605 and ZTFJ053450+212603. 


To search for periodic signals, we applied the Lomb-Scargle algorithm \citep{1982ApJ...263..835S} to the forced photometry of all 202 light curves using \texttt{gatspy} \citep{2015zndo.....14833V}. We searched periods of 2 minutes to 10 days. We quantified the significance of a detected period by dividing the Lomb-Scargle power of the best-fit period by the median Lomb-Scargle power of all searched periods. We found three sources with periods significant in the 90th percentile.



\begin{deluxetable*}{c|c|c|c|c}
 \tablehead{
 \colhead{Name} & \colhead{RA (J2000)} & \colhead{DEC (J2000)} & \colhead{\parbox{5cm}{\vspace{5pt}\centering Distance from \textit{NuSTAR} source (arcsec)}} & \colhead{Notes}
 }
 \tablecaption{Candidate Counterparts to \textit{NuSTAR} J053449+2126.0}
 \startdata
 2MASS J05344864+2125080 & 05 34 48.644  & +21 25 8.04 & 55.4 & Palomar Gattini-IR detection \\
 ZTFJ053452+212605 & 05 34 52.004  & +21 26 5.74 & 39.3 & ZTF $>2$ mag burst \\
 ZTFJ053450+212603 & 05 34 50.700  & +21 26 3.44 & 21.0 & ZTF $>2$ mag burst \\
 ZTFJ053450+212634 & 05 34 50.264  & +21 26 34.18 & 34.6 & ZTF periodicity \\
 ZTFJ053449+212632 & 05 34 49.888  & +21 26 32.22 & 30.8 & ZTF periodicity \\
 ZTFJ053447+212605 & 05 34 47.499  & +21 26 5.78 & 23.9 & ZTF periodicity\\
\enddata
{{\centering No obvious optical or infrared counterpart to \textit{NuSTAR} J053449+2126.0 was identified. The above objects are identified as the most likely candidates based on outbursts or periodicity seen in their light curves.}}
\end{deluxetable*}

\label{tab:all_sources}

\section{Analysis of Potential  
Candidates and Other Constraints}
All 6 candidates (summarized in Table \ref{tab:all_sources}) were queried on CDS Portal and all featured unremarkable spectral energy distributions (SEDs) with the possible exception of ZTFJ053447+212605, which features a slight infrared excess in its SED. SED data was compiled from \textit{Gaia}, SDSS, Pan-STARRS, UKIDSS, and WISE photometry. Catalysmic variables are Galactic X-ray emitters which may also have infrared excesses \citep{2009ASPC..404..234H}. This would then attribute \textit{NuSTAR} J053449+2126.0 to the X-ray counterpart to a CV. Both non-magnetic and magnetic CVs can have X-ray luminosities $L_X \sim 10^{31}$ erg/s \citep{2017PASP..129f2001M}, but we did not find a CV orbital period in the data. 

None of the candidates stand out as an obvious counterpart to a hard X-ray source. The candidates identified through their long-term light curve behavior do not show sustained or exceptionally high-amplitude outbursts. The candidates identified through periodicity do not show highly significant periods (i.e. $>$ 95th percentile) and feature variation on the order of the error bars of each photometric data point.


Since \textit{NuSTAR} J053449+2126.0 is located close to the Crab nebula in the sky, we suspect it is a Galactic source. The extinction to this source is obtained from the NASA/IPAC Extragalactic Database, which derives its results from \cite{2011ApJ...737..103S}. We find the extinction in the direction of \textit{NuSTAR} J053449+2126.0 to be $A_V \sim 1.37$.

Galactic sources that would have shown strong optical periodicity at high $>$1 mag amplitudes would be attributed to ``spider variables". These are systems where a pulsar irradiates a companion low-mass star or brown dwarf and the resulting shock emits in X-rays. Cyclotron beaming from magnetic CVs known as polars also lead to high-amplitude periodic variability. Since no folded light curves with amplitudes $>$1 mag were found in the ZTF archive, we can exclude polar CVs and spider variables from the possible Galactic associations with \textit{NuSTAR} J053449+2126.0.

\section{Discussion and Conclusion}

X-ray imaging with a telescope with a sharper point spread function such as \textit{Swift} or \textit{Chandra} is needed. This would provide better spatial resolution of the \textit{NuSTAR} J053449+2126.0 1 arcmin region as well as provide the soft X-ray coverage when comparing to the \textit{NuSTAR} spectrum. Additional X-ray data would also inform us if \textit{NuSTAR} J053449+2126.0 is a persistent or transient source.

We have found no obvious counterpart to \textit{NuSTAR} J053449+2126.0 after searching ZTF and Palomar Gattini-IR alerts as well as long-term photometry. Our limited knowledge of the temporal behavior of \textit{NuSTAR} J053449+2126.0 motivates the need for further observations within the 1 arcmin region to determine the nature of the source.

\bibliography{main}{}
\bibliographystyle{aasjournal}

\end{document}